# Quantum-size effect and tunneling magnetoresistance in ferromagnetic-semiconductor quantum heterostructures

Shinobu Ohya[1], Pham Nam Hai[1], Yosuke Mizuno[1] and Masaaki Tanaka[1,2]

1. Department of Electronic Engineering, The University of Tokyo,
7-3-1 Hongo, Bunkyo-ku, Tokyo 113-8656, Japan

2. Japan Science and Technology Corporation,
4-1-8 Honcho, Kawaguchi, Saitama 332-0012, Japan

**Abstract**

We report on the resonant tunneling effect and the increase of tunneling magnetoresistance (TMR) induced by it in ferromagnetic-semiconductor GaMnAs quantum-well heterostructures. The observed quantum levels of the GaMnAs quantum well were successfully explained by the valence-band $k \cdot p$ model with the $p$-$d$ exchange interaction. It was also found that the Fermi level of the electrode injecting carriers is important to observe resonant tunneling in this system.

III-V-based ferromagnetic-semiconductor heterostructures containing GaMnAs are hopeful candidates for future spintronic devices. Large tunneling magnetoresistance (TMR) of 75% (8.0 K) [1] and 290% (0.39 K) [2] were observed in the GaMnAs-based single-barrier heterostructures. Also, the GaMnAs-based quantum heterostructures are expected to realize new functions by combining a spin degree of freedom and the quantum-size effect (resonant tunneling effect) [3 - 11]. For example, a large enhancement of TMR up to ~800%[4] and to more than ~$10^6$%[3] is expected in GaMnAs-based resonant tunneling diode (RTD) structures. In magneto-optical measurements of the GaMnAs quantum well (QW), blue shifts of the magneto-optical spectra have been observed, suggesting the existence of the quantum-size effect in GaMnAs[12,13]. However, there are no reports on clear observation of the resonant tunneling effect in ferromagnetic-semiconductor quantum heterostructures[3,14,15]. On the other hand, in the metal-based magnetic tunnel junctions (MTJs), the resonant tunneling effect and the TMR oscillation induced by it have been observed[16,17]. Comparing with these metal systems, the GaMnAs-based quantum heterostructures are fully epitaxial single crystals and have atomically flat interfaces, where highly coherent tunneling is expected. Therefore, this is a material system ideal for physical understanding of the spin-dependent resonant tunneling effect. Also it has good compatibility with III-V-based semiconductor devices, thus can lead to future three-terminal spin devices such as resonant tunneling spin transistors. Therefore, it is a very important issue to observe resonant tunneling effect in III-V-based ferromagnetic heterostructures. In this Letter, we report a first successful detection of resonant tunneling effect and TMR increase induced by it in semiconductor heterostructures.

The GaMnAs-QW double-barrier quantum heterostructures examined here were grown by molecular beam epitaxy (MBE) with a layer structure from the surface side; $Ga_{0.95}Mn_{0.05}As$(20 nm)/ GaAs(1 nm)/ $Al_{0.5}Ga_{0.5}As$(4 nm)/ GaAs(1 nm)/ $Ga_{0.95}Mn_{0.05}As$($d$ nm)/ GaAs(1 nm)/ AlAs(4 nm)/ Be-doped GaAs(100 nm) on a $p$-type GaAs(001) substrate, where $d$ ranges from 3.8 to 20 nm. The Be concentration of the Be-doped GaAs (GaAs:Be) layer was $1\times10^{18}$ cm$^{-3}$. The 1-nm thick GaAs layers were inserted to prevent the Mn diffusion into the barrier layers and to smooth the surface. The GaAs:Be, AlAs and the lowest GaAs spacer layers were grown at 600, 550, and 600 ºC, respectively. The GaMnAs layers were grown at 225 ºC, and GaAs/ AlGaAs/ GaAs layers below the top GaMnAs layer were grown at 205 ºC. We grew four samples named A, B, C, and D. When growing sample A, we moved the in-plane position of the main shutter equipped in our MBE chamber in front of the sample surface during the growth of the GaMnAs QW, and changed $d$ from 3.8 to 8.0 nm on the same sample wafer. In the growth of samples B, C, and D, $d$ was fixed at 12, 16, and 20 nm, respectively. After the growth, circular mesa diodes with 200 μm in diameter were fabricated by chemical etching. We spin-coated an insulating negative resist on the sample, opened a contact hole with 180 μm in diameter on the top of the mesa, and fabricated a metal electrode by evaporating Au on this surface. In the following measurements, the bias polarity is defined by the voltage of the top GaMnAs electrode with respect to the substrate. The schematic band diagrams when negative and positive biases $V$ are applied to this structure are shown in Fig. 1(a) and 1(b), respectively. In this structure, TMR occurs by tunneling of holes between the ferromagnetic GaMnAs top electrode and the ferromagnetic GaMnAs QW. The following tunneling transport measurements were carried out in a cryostat cooled at 2.6 K with a conventional two-terminal direct-current (DC) method. $dI/dV$-$V$ and $d^2I/dV^2$-$V$ characteristics were derived mathematically from the data of the $I$-$V$ characteristics measured at every 5 mV. The results of the bias dependence of TMR were obtained mathematically from the data of $I$-$V$ characteristics measured in parallel and anti-parallel magnetizations. $I$-$V$ characteristics in both parallel and anti-parallel magnetizations were obtained at zero-magnetic field.

Figure 1(c) shows $d^2I/dV^2$-$V$ characteristics of these junctions in parallel magnetization at 2.6 K. Sharp features near the zero bias are observed in all the curves with various $d$, corresponding to the zero-bias anomaly which is usually observed in GaMnAs-based heterostructures[18]. The most important feature in Fig. 1(c) is the oscillations whose peak voltages depend on $d$ in the negative bias region of all the curves. With increasing $d$, these peaks shift to a smaller voltage and the period of the oscillation becomes short. Such oscillatory behaviors have not been observed in GaMnAs-based single-barrier MTJs, indicating that these oscillations are induced by the resonant tunneling effect. The peaks HH$n$ and LH$n$ ($n$=1, 2, 3 ⋯) are assigned to resonant tunneling through the $n$th level of the heavy hole (HH) band and light hole (LH) band in the GaMnAs QW, respectively, as will be described later. Figure 1(d) shows the $dI/dV$-$V$ curves of the junction with $d$=12 nm at 2.6 K in parallel (blue curve) and antiparallel (red curve) magnetization. These two curves are almost the same, but there is a little voltage shift which is most obvious at LH1.

    TMR was clearly observed in the junctions with $d$ from 3.8 to 12 nm. Figure 2 shows the bias dependence of TMR of these junctions at 2.6 K with a magnetic field applied in plane along the [100] direction. Here, the TMR ratio is defined as ($R_{AP}$-$R_P$)/$R_P$, where $R_P$ and $R_{AP}$ are the tunnel resistances (=$V/I$) for parallel and antiparallel magnetization, respectively. The TMR values are normalized by the maximum value in each curve. The inset is the magnetic field dependence of the tunnel resistance obtained in the junction with $d$=12 nm when the bias voltage is +10 mV (red curve) and -104 mV (blue curve), exhibiting typical TMR curves with TMR ratios of 18.9% and 14.1%, respectively. In the main graph of Fig. 2, TMR oscillations can be seen in the negative bias region of all the curves. With increasing $d$, the TMR peaks (except for that near zero bias) shift to a smaller voltage as is the case of the $d^2I/dV^2$-$V$ characteristics shown in Fig. 1(c), indicating that these TMR increases are induced by the resonant tunneling effect. Especially, a large TMR increase occurs at LH1, which is caused by the magnetization-dependent peak's shift at LH1 observed in the $dI/dV$-$V$ characteristics shown in Fig. 1(d). Also, Fig. 2 indicates that the bias voltage of $V_{half}$, at which TMR is reduced by half, can be significantly increased by the resonant tunneling effect. In the negative-bias region of $d$=12 nm, |$V_{half}$| was increased to 124 mV due to the TMR increase at LH1, whereas $V_{half}$ in the positive bias was 76 mV. This value is much higher than those of GaMnAs-based single-barrier structures which are usually around 50 mV [2,19,20]. On the other hand, the origin of the unusual sharp decrease of TMR near the zero bias is not clear, but it is probably partly due to the unknown band offset of the GaMnAs QW mentioned below.

    To understand the experimental results, we calculated the quantum levels of GaAs/ Al$_{0.5}$Ga$_{0.5}$As(4 nm)/ GaMnAs($d$ nm)/ AlAs(4 nm)/ GaAs, using the transfer matrix method[21] with the valence-band 4×4 $k$·$p$ Hamiltonian[22] and the $p$-$d$ exchange Hamiltonian[23] for including the in-plane magnetization of the GaMnAs QW. In this calculation, we assumed that the splitting energy $\Delta$, which corresponds to the spin splitting energy for the LH valence band at the Γ point, is (3 meV, 0, 0) along the

in-plane [100] direction parallel to the magnetic field applied in our experiments. Figure 3 shows the calculated result of the resonant-peak bias voltages for the heavy hole tunneling (red points and curves assigned to HH$n$) and the light hole tunneling (blue points and curves assigned to LH$n$) at $k_\parallel$=0, where $k_\parallel$ is the wave vector parallel to the film plane. These resonant peak bias voltages were derived by multiplying the calculated energy values of the quantum levels by 2.5 (ideally this value is 2.) for better fit, which means that 20% of the applied bias voltage is consumed in the electrodes. The center energy of the spin-split valence band of the GaMnAs QW was set at 28 meV above the valence band of GaAs in terms of hole energy at the Γ point. The origin of this band offset introduced here is not clear at present. Since the heavy hole spins are oriented along the tunneling direction and the *p-d* exchange Hamiltonian is proportional to **s·S**, the quantum levels of HH are not spin split but those of LH are split by the in-plane magnetization[24], where **s** and **S** are spins of the carrier and the Mn atom, respectively. Holes tunnel through the lower and upper states of the spin-split LH quantum levels in parallel and in antiparallel magnetization, respectively. The calculated voltages of the LH quantum levels shown in Fig. 3 are those in parallel magnetization.

Figure 3 also shows the experimental peak voltages assigned to HH and LH quantum levels observed in the $d^2I/dV^2$-$V$ curves in parallel magnetization by black solid rectangles and triangles, respectively. Here, the peak voltages are expressed in the absolute values. Although there is a little deviation between the experimental and calculated results, the experimental results are well fitted by the present model. Possible origins of these deviations are considered as follows. When *d* is thinner than 5 nm, Mn diffusion from the GaMnAs QW layer to the adjacent GaAs spacer layer yields lowering of the resonant tunneling energy. The deviations may also come from the point defects incorporated into GaMnAs such as Mn interstitials[25] and As anti-site defects[26], whose concentrations are very sensitive to the growth condition of GaMnAs[27]. These defects can influence the spin-splitting energy, band offset, and strain, leading to such deviations. We note that the spin-splitting energy (3 meV, 0, 0) is consistent with our results of the temperature dependence of TMR, where TMR disappeared at around 20 K when increasing temperature[28,29]. Also, this spin-splitting energy can well explain the quantity of the peak voltage shift of 7(±2) mV observed in the *dI/dV*-*V* curves with parallel and antiparallel magnetizations at LH1 shown in Fig. 1(d) (*i.e.* 7 ≈ 3 × 2.5). If a GaMnAs QW with a larger spin-splitting energy can be obtained by using a better growth condition, a larger TMR enhancement will be obtained. Our result that the quantum levels of the GaMnAs QW can be described by the spin-split valence-band model indicates that the tunneling phenomena through GaMnAs are strongly associated with the valence-band feature.

Finally, we briefly discuss the reason why the resonant peaks were observed only in the negative bias region when holes are injected from the GaAs:Be layer to the GaMnAs QW. We think that it is attributed to the difference of the Fermi energy between the GaMnAs top and the GaAs:Be bottom electrodes[28]. GaMnAs has a large hole concentration of the order of $10^{20}$-$10^{21}$ cm$^{-3}$, thus a large Fermi level around 200 meV. This indicates that the holes occupy a wide $k_{||}$ region. Figure 4(a) shows the hole subband structure of AlAs(1 nm)/ GaAs QW(5 nm)/ AlAs(1 nm) calculated by the *k·p* model and energy regions corresponding to these subbands when carriers are injected from such an electrode with a large Fermi level. The gray region is the $k_{||}$ region to which holes can tunnel, and the black bands are the corresponding energy regions. (In Fig. 4, we ignored the spin splitting and used a GaAs QW and an AlAs barrier instead of GaMnAs and AlGaAs, respectively, for simplicity.) In this case of Fig. 4(a), these energy regions of the subbands are energetically overlapped, thus these subbands cannot be detected separately in tunneling transport measurements. On the other hand, if carriers are injected from GaAs:Be with a smaller carrier concentration of $1\times10^{18}$ cm$^{-3}$ whose Fermi level is estimated to be about 7 mV, the tunneling occurs only within a small $k_{||}$ region. Figure 4(b) shows the subbands and corresponding energy regions when carriers are injected from such an electrode with a small Fermi level, where these energy regions are energetically separated. Therefore, the resonant peaks were observed only in the negative bias region. Our result indicates that electrodes with a low carrier concentration are appropriate for clear detection of the resonant tunneling effect in GaMnAs QW heterostructures.

In summary, we have observed the resonant tunneling effect and TMR increase induced by it in GaMnAs-QW double-barrier heterostructures. The observed quantum levels of GaMnAs QW were successfully explained by the valence-band *k·p* model with the *p-d* exchange interaction. It was also found that electrodes with a low carrier concentration are appropriate for clear detection of the resonant tunneling effect in this system.

This work was partly supported by SORST of JST, Grant-in-Aids for Scientific Research, IT Program of RR2002 of MEXT, and Kurata-Memorial Hitachi Science & Technology Foundation.

**References**

[1] M. Tanaka and Y. Higo, Phys. Rev. Lett. **87,** 026602 (2001).
[2] D. Chiba, F. Matsukura, and H. Ohno, Physica E **21,** 966 (2004).
[3] T. Hayashi, M. Tanaka, and A. Asamitsu, J. Appl. Phys. **87,** 4673 (2000).
[4] A. G. Petukhov, A. N. Chantis, and D. O. Demchenko, Phys. Rev. Lett. **89,** 107205 (2002).
[5] S. S. Makler, M. A. Boselli, J. Weberszpil, X. F. Wang, and I. C. da Cunha Lima, Physica B **320,** 396 (2002).

[6] F. Giazotto, F. Taddei, R. Fazio, and F. Beltram, Appl. Phys. Lett. **82,** 2449 (2003).
[7] T. Uemura, T. Marukame, and M. Yamamoto, IEEE Trans. Magn. **39,** 2809 (2003).
[8] I. Vurgaftman and J. R. Meyer, Appl. Phys. Lett. **82**, 2296 (2003).
[9] H. –B. Wu, K. Chang, J. –B. Xia, and F. M. Peeters, J. Supercond. **16**, 279 (2003).
[10] N. Lebedeva and P. Kuivalainen, P. Phys. Stat. Sol. (b) **242**, 1660 (2005).
[11] S. Ganguly, L. F. Register, S. Banerjee, and A. H. MacDonald, Phys. Rev. B **71,** 245306 (2006).
[12] H. Shimizu and M. Tanaka, J. Appl. Phys. **91,** 7487 (2002).
[13] A. Oiwa, R. Moriya, Y. Kashimura, and H. Munekata, J. Magn. Magn. Mater. **272-276,** 2016 (2004).
[14] R. Mattana *et al.,* Phys. Rev. Lett. **90,** 166601 (2003).
[15] S. Ohya, P. N. Hai, and M. Tanaka, Appl. Phys. Lett. **87,** 012105 (2005).
[16] T. Nozaki, N. Tezuka, and K. Inomata, Phys. Rev. Lett. **96,** 027208 (2006).
[17] S. Yuasa, T. Nagahama, and Y. Suzuki, Science **297,** 234 (2002).
[18] S. H. Chun, S. J. Potashnik, K. C. Ku, P. Schiffer, and N. Samarth, Phys. Rev. B **66,** 100408(R) (2002).
[19] R. Mattana *et al.,* Phys. Rev. B, **71**, 075206 (2005).
[20] M. Elsen *et al.,* Phys. Rev. B **73,** 035303 (2006).
[21] R. Wessel and M. Altarelli, Phys. Rev. B **39**, 12802 (1989).
[22] J. M. Luttinger and W. Kohn, Phys. Rev. B **97,** 869 (1955).
[23] T. Dietl, H. Ohno, and F. Matsukura, Phys. Rev. B **63,** 195205 (2001).
[24] M. Sawicki *et al.*, Phys. Rev. B **70,** 245325 (2004).
[25] K. M. Yu *et al.,* Phys. Rev. B **65,** 201303(R) (2002).
[26] B. Grandidier *et al.,* Appl. Phys. Lett. **77,** 4001-4003 (2000).
[27] H. Shimizu, T. Hayashi, T. Nishinaga, and M. Tanaka, Appl. Phys. Lett. **74,** 398 (1999).
[28] H. Ohno *et al.,* Appl. Phys. Lett. **73,** 363-365 (1998).
[29] In ref. 28, it was reported that the spin-splitting energy of the HH valence band of GaMnAs with a Curie temperature of ~70 K was estimated to be 44 meV. If we simply assume the spin-splitting energy of HH is three times larger than that of LH, the spin-splitting energy of the LH valence band in their sample is ~15 meV. The Curie temperature of the GaMnAs layers in our heterostructures is estimated to be around 20 K, thus the value of the spin-splitting energy of 3 meV of our GaMnAs layers is reasonable (*i.e.* $3 \approx 15 \times 20/70$).

**Figure Captions**

**FIG. 1**  (a), (b) Schematic band diagrams of the $Ga_{0.95}Mn_{0.05}As$(20 nm)/ GaAs(1 nm)/ $Al_{0.5}Ga_{0.5}As$(4 nm)/ GaAs(1 nm)/ $Ga_{0.95}Mn_{0.05}As$(d nm)/ GaAs(1 nm)/ AlAs(4 nm)/ GaAs:Be (100 nm) RTD junction when the bias polarity is negative and positive, respectively.  Here, the 1-nm-thick GaAs spacer layers are omitted for simplicity. (c) $d^2I/dV^2$-$V$ characteristics of these RTD junctions with various QW thicknesses $d$ in parallel magnetization at 2.6 K.  Numbers in the parentheses express the magnification ratio for the vertical axis.  (d) $dI/dV$-$V$ characteristics of the junction with $d$=12 nm at 2.6 K in parallel (blue curve) and antiparallel (red curve) magnetization.

**FIG. 2**  Bias dependence of TMR in $Ga_{0.95}Mn_{0.05}As$(20 nm)/ GaAs(1 nm)/ $Al_{0.5}Ga_{0.5}As$(4 nm)/ GaAs(1 nm)/ $Ga_{0.95}Mn_{0.05}As$(d nm)/ GaAs(1 nm)/ AlAs(4 nm)/ GaAs:Be(100 nm) RTD junctions with various QW thicknesses $d$ when a magnetic field was applied in plane along the [100] direction at 2.6 K, where the TMR ratios are normalized by the maximum value of TMR in each curve.  Here, the TMR ratio is defined as $(R_{AP}-R_P)/R_P$, where $R_P$ and $R_{AP}$ are the tunnel resistances (=$V/I$) for parallel and anti-parallel magnetization, respectively.  These results were obtained at zero-magnetic field.  The inset is the magnetic field dependence of the tunnel resistance obtained in the junction with $d$=12 nm when the bias voltage is +10 mV (red curve) and -104 mV (blue curve), showing typical TMR curves with TMR ratios of 18.9% and 14.1%, respectively.

**FIG. 3**  Calculated and experimentally obtained resonant peak voltage *vs.* the GaMnAs QW thickness $d$.  The black solid rectangles and triangles denote the experimentally obtained resonant-peak voltages assigned as HH and LH quantum levels in the $d^2I/dV^2$-$V$ characteristics of $Ga_{0.95}Mn_{0.05}As$(20 nm)/ GaAs(1 nm)/ $Al_{0.5}Ga_{0.5}As$(4 nm)/ GaAs(1 nm)/ $Ga_{0.95}Mn_{0.05}As$(d nm)/ GaAs(1 nm)/ AlAs(4 nm)/ GaAs:Be(100 nm) RTD junctions in parallel magnetization, respectively.  Here, these voltages are expressed in the absolute values.  The small red and blue points (curves) denote the calculated resonant voltages of HH and LH, respectively.  The quantum levels of LH1 and LH2 are spin-split by the in-plane magnetization introduced by the *p-d* exchange Hamiltonian.  The calculated voltages of the LH quantum levels shown in this figure are those in parallel magnetization.

**FIG. 4**  Hole subband structure of AlAs(1 nm)/ GaAs QW (5 nm)/ AlAs(1 nm)

calculated by the *k·p* model and the energy regions corresponding to these subbands when holes are injected from *p*-type GaAs (a) with a large Fermi level of about 200 mV and (b) with a small Fermi level of about 7 meV.   The gray regions correspond to the $k_\parallel$ regions where tunneling can occur, and the black bands are corresponding energy regions.   The broken lines are maximum and minimum energies of each subband within the gray region.   Here, we ignored the spin splitting for simplicity.

**FIG. 1**

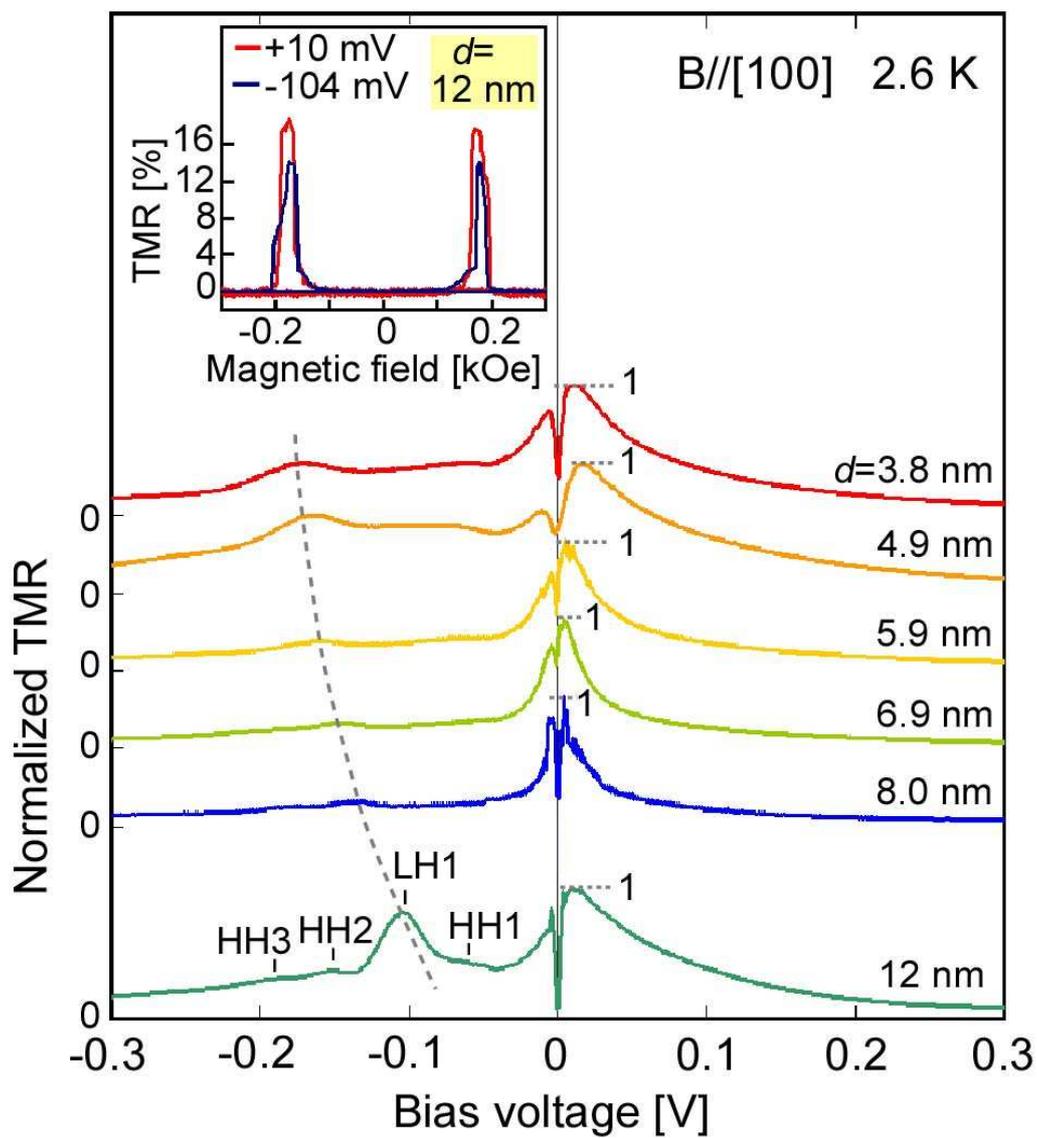

**FIG. 2**

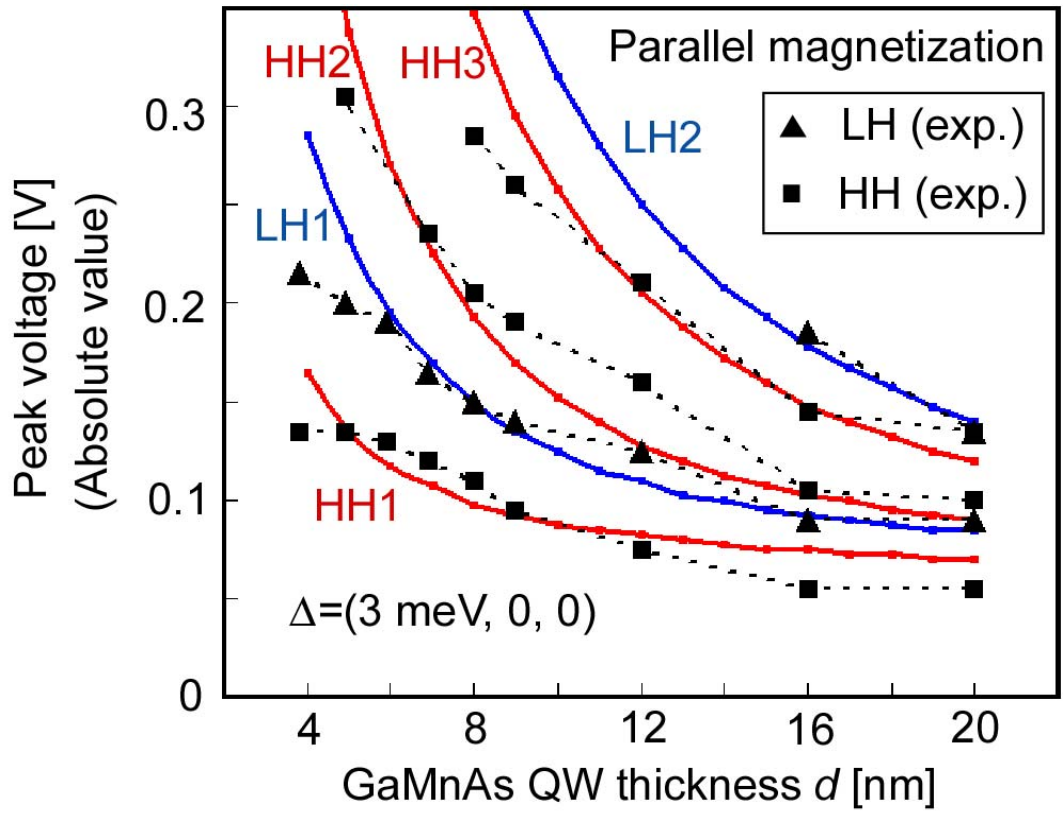

**FIG. 3**

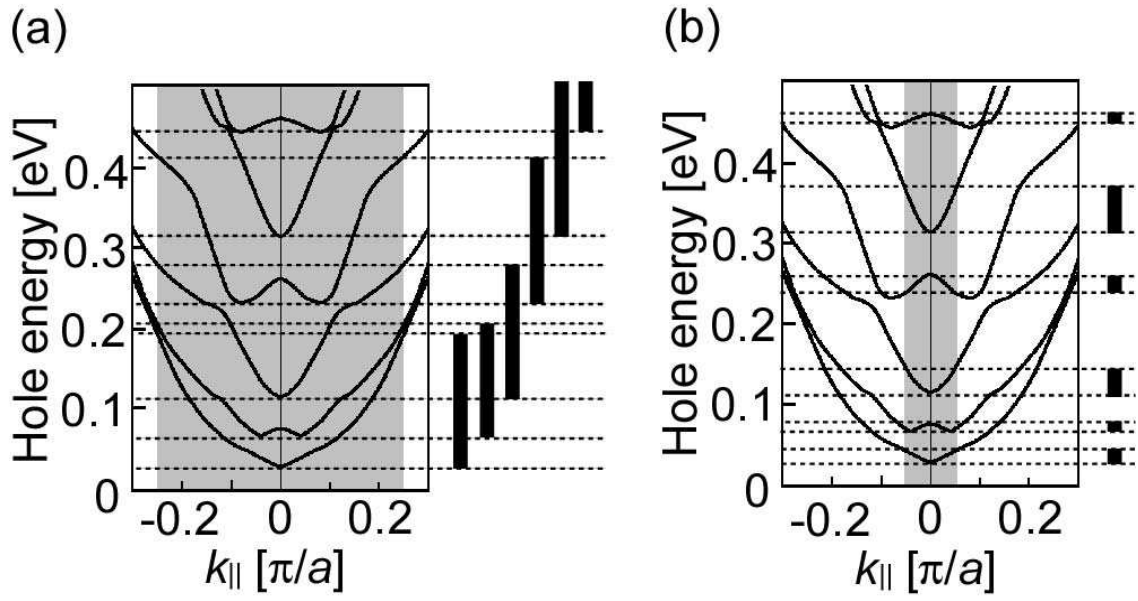

**FIG. 4**